\documentclass[12pt]{article}
\usepackage{xcolor}
\usepackage{multicol}
\usepackage{graphicx}
\usepackage{amsmath}
\usepackage{amsthm}
\usepackage{amssymb}
\usepackage[english]{babel}
\usepackage[left=1in,right=1in,bottom=1in,top=1in,a4paper]{geometry}
	\newtheorem{defi}{Definition}
\usepackage{setspace}
\usepackage{maplestd2e}

\title{Computation of generating symmetries}
\author{Alexander G. Rasin  \\
Department of Mathematics,\\ 
Ariel University, Ariel 40700, Israel}

\begin{document}
\maketitle

%\begin{frontmatter}
%\title{Computation of generating symmetries}
%\author{Alexander G. Rasin}

%\affiliation{organization={Department of Mathematics, Ariel University},%Department and Organization
%                     city={Ariel},
%            postcode={40700}, 
%                      country={Israel}}

\begin{abstract}
In this article we continue to develop the theory of generating symmetries for integrable equations. A technique for computation of generating symmetries using Maple is presented. The technique is based on the standard symmetry method. By using it we find generating symmetries for the KdV, Camassa-Holm, mKdV, sine-Gordon, Boussinesq, associated Degasperis-Procesi and associated Novikov equations.
\end{abstract}
%\end{frontmatter}	
\section{Introduction}
The study of symmetries of differential equations goes back to Sophus Lie. The technique for computing the point symmetry group is well described in \cite{olverbook, hydonbook,blumanbook}. At first, symmetry analysis was applied to the particular PDEs in order to understand the geometry of equations and to generate group-invariant solutions \cite{f2013}. With the development of computer algebra more complicated tasks became reachable. Among them are the classification of point symmetry groups for classes of PDEs \cite{p2004new,i1995crc,v2009e}, symmetry analysis of difference equations \cite{HR,HYb} and the derivation of high-order and non-local symmetries \cite{AGI,VKS}.

Integrable differential equations typically display substantially more symmetry properties than their non-integrable
counterparts (with the exception of linear systems, which also may be regarded as integrable). It was found that many integrable equations have infinite hierarchies of local and non-local symmetries \cite{Lou,GH}. These hierarchies can be obtained via recursion operators \cite{Ol1,olverbook,lenard}, master-symmetries \cite{f1983m}, generating symmetries \cite{RSG} and so on. The implementation of known techniques for particular equations is usually an art rather then routine computing. 

In \cite{RSG} we introduced generating symmetries (GSs). This kind of symmetry is an infinitesimal transformation which depends upon the solution of the equation and the function appearing in a B\"acklund transformation or in a Lax pair. The characteristic of the GS is a generating function for the entire hierarchy of infinitesimal symmetries of the corresponding PDE. That is, when expanded in a suitable series of powers of the spectral parameter, all the components are symmetries.
In \cite{RSG} GSs were found for KdV, associated Camassa-Holm and sine-Gordon equations and later this result was extended for other equations \cite{RS3,RS4,RS5,RS6}. A major advance in the research of GSs was achieved in \cite{RS7}. \textit{All} the symmetries of KdV were expressed there using four GSs. This was done by introducing a new GS, which gives a hierarchy of symmetries with explicit dependence upon $x$ and $t$. The commutation relations of GSs which describe the algebras of all symmetries, including non-local ones, were given in \cite{RS7}, as well. 

Apart from generating infinite hierarchies of symmetries, GSs can also be used to obtain new solutions. Since GSs depend on functions that appear in a B\"acklund transformation or Lax pair, they are nonlocal, and there are few existing results on the use of nonlocal symmetries to generate solutions. An exception, though, is in the papers \cite{Loub,lou2012}. There the authors use some of the GSs of the KdV equation to generate a variety of interesting solutions, for example solitons and dark solitons residing on a cnoidal wave background (formulas (44),(47) in \cite{Loub}). Two methods are used in \cite{Loub,lou2012}: finding finite transformations associated with the GSs, and a generalization of the standard similarity reduction technique. The first is possible as some GSs are infinitesimal forms of superpositions of B\"acklund transformations \cite{RSG,RS3,RS4,RS5,RS6}. The second requires finding the action of GSs on the functions that appear in the B\"acklund transformation or Lax pair. This is discussed at length in \cite{RS7}.

The generating symmetries in \cite{RSG,RS3,RS4,RS5,RS6} were obtained via a limit process from a superposition principle. This approach has the following disadvantages: a) it is necessary to know the superposition principle (if it exists); b) not all superposition principles allow this procedure; c) not all GSs can be found. These difficulties motivated us to find better approach to compute GSs.

In this article we propose a direct method for computation of GSs. This method is an upgrade of the standard symmetry method described in \cite{olverbook,Ovs}. We will call it the generating symmetry method (GSM). GSM requires the existence of a Lax pair or a B\"acklund transformations for the corresponding equation. This requirement is fulfilled for integrable equations, since this is one of the main integrability properties. The main advantages of GSM are the possibility of application to a vast class of equations and a very simple implementation, which can be done via computer algebra. 

The structure of this paper is as follows: In section 2, GSM is presented.
In section 3, GSM is applied to equations with a second order Lax pair. In section 4, GSM is applied to equations with a third order Lax pair.
Section 5 contains some concluding comments and questions for further study. In the Appendix we show the Maple program for computation of GSs.
\section{The generating symmetry method}
The idea of the method is presented here for a scalar partial differential equation for a single function of two independent variables, $u=u(x,t)$. It can be extended for a system of equations with many variables.  

By a scalar PDE we mean 
\begin{equation}F(u,u_x,u_t,u_{xx},u_{tx},u_{tt},...)=0,\label{eq1}\end{equation}
where $u_x, u_t,...$ are derivatives of $u$.
The infinitesimal generator for a symmetry of (\ref{eq1}) is a vector field 
\[
X=\eta\frac{\partial}{\partial u}.\label{X1}
\]
Here $\eta$ satisfies the linearisation of (\ref{eq1})  
\begin{equation}
\eta F_u+\eta_xF_{u_x}+\eta_tF_{u_t}+\eta_{xx}F_{u_{xx}}+...=0.\label{LEQ}
\end{equation}
$\eta$ is called the characteristic of $X$ and it depends upon $x,t,u$ and the derivatives of $u$.

Let us assume that (\ref{eq1}) has a Lax pair. A Lax pair is a system of linear differential equations of the form
\begin{eqnarray}
L\psi&=&\lambda\psi,\label{LP1}\\
\psi_t&=&P\psi,\label{LP2}
\end{eqnarray}
where $L,P$ are differential operators and $\lambda$ is the spectral parameter. Solutions of the Lax pair are called eigenfunctions. 
A Lax pair has several (depending upon the order of the equation (\ref{LP1})) linearly independent solutions
\begin{equation}
\psi^{(1)}, \psi^{(2)},... .
\end{equation}
Often a Lax pair can be transformed by the substitution 
\begin{equation}
v=\frac{\psi_x}{\psi}\label{VLP}
\end{equation}
to the system which defines a B\"acklund transformation. 

GSs were introduced in \cite{RSG, RS3, RS4}. In these articles the characteristics $\eta$ of the GSs depend upon solutions of B\"acklund transformations and their derivatives with respect to $x$
\begin{equation}
v^{(1)}, v^{(2)},v^{(1)}_x, v^{(2)}_x.\label{var1}
\end{equation} 
In \cite{RS7} we show a GS which also depends upon the derivative of solutions of B\"acklund transformations with respect to the spectral parameter 
\begin{equation}
v^{(1)}_{\lambda}, v^{(2)}_{\lambda}.\label{var2}
\end{equation}  
So, in order to find general GSs the variables (\ref{var1},\ref{var2}) have to be included in $\eta$.  
Equivalently, the solutions of the Lax pair can be used. In this case $\eta$ should depend on 
\[\psi^{(1)}, \psi^{(2)}, \psi^{(1)}_x, \psi^{(2)}_x, \psi^{(1)}_{\lambda}, \psi^{(2)}_{\lambda},... .\] 
\begin{defi} A vector field (\ref{X1}) is a \textbf{generating symmetry} of equation (\ref{eq1}), if $\eta$ depends upon the variables of a Lax pair or a B\"acklund transformation and if (\ref{LEQ}) holds on solutions of (\ref{eq1}) and solutions of the Lax pair or the B\"acklund transformation.
\end{defi}
The GSM is based on the standard symmetry method. Namely, (\ref{LEQ}) should be satisfied on solutions of (\ref{eq1},\ref{LP1},\ref{LP2}). The technical details of the method can be found in the Appendix where we present the Maple program for computation of GSs for the potential KdV equation.
\section{Generating symmetries for equations with a second order Lax pair}
The general form of (\ref{LP1}) in the second order Lax pair is 
\begin{equation}
\psi_{xx}=F_1\psi_x+F_2\psi.\label{LPt1}
\end{equation}
Here $F_1, F_2$ are functions which depend upon the parameter $\lambda$, variables $t,x,u$ (see \cite{Steeb} for an example of the Lax pair with explicit dependence upon $t$ and $x$) and the derivatives of $u$.

\noindent{\bf Notes:} 
\begin{enumerate}
	\item  The second order Lax pair has two linearly independent solutions: $\psi^{(1)}$ and $\psi^{(2)}$.
	\item  The two solutions $\psi^{(1)}, \psi^{(2)}$ satisfy 
\begin{equation}
W(\psi^{(1)},\psi^{(2)})=\psi^{(1)}\psi^{(2)}_x-\psi^{(1)}_x\psi^{(2)}=ce^{\int F_1 dx},\label{pp1}
\end{equation}
where $W$ is the Wronskian and $c$ is a constant of integration independent of $x$. The dependence of $c$ upon $t$ can be determined from the consistency of (\ref{pp1}) and the Lax pair. We assume that $\int F_1 dx$ is local. 
\end{enumerate}
In order to find generating symmetries for an equation with a second order Lax pair we take $\eta$ in the following form
\begin{equation}
\eta=\eta(x,t,u,u_x,u_t,\psi^{(1)}, \psi^{(2)},\psi^{(1)}_x, \psi^{(1)}_{\lambda}, \psi^{(2)}_{\lambda}).\label{etapk}
\end{equation}
$\eta$ doesn't depend upon $\psi^{(2)}_x$, since it can be eliminated with the help of (\ref{pp1}). $\eta$ depends upon $\lambda$, but we don't include it in (\ref{etapk}), since it can be considered as a constant.
\subsection{The KdV equation}
The KdV equation is
\begin{equation}
  \phi_t - 3\phi\phi_x - \textstyle{\frac14} \phi_{xxx} = 0 \ . \label{KdV} 
\end{equation}  
It is connected to the potential KdV equation (pKdV) via $\phi=u_x$
\begin{equation}
  u_t - \textstyle{\frac32} u_x^2 - \textstyle{\frac14} u_{xxx} = 0 \ . \label{pKdV} 
\end{equation}  
The Lax pair for KdV and pKdV \cite{cn1} is
\begin{eqnarray}
\psi_{xx}&=&(\lambda-2u_x)\psi,\label{LPkdv1}\\
\psi_t&=&(\lambda+u_x)\psi_x-\textstyle{\frac12} u_{xx}\psi.\label{LPkdv2}
\end{eqnarray}
The coefficient of $\psi_x$ is zero, therefore from (\ref{pp1}) we get
\begin{equation}
\psi^{(2)}_x=\frac{\psi^{(1)}_x\psi^{(2)}+c}{\psi^{(1)}}.\label{plk}
\end{equation}
By checking the consistency with (\ref{LPkdv1},\ref{LPkdv2}) we obtain that $c$ does not depend upon $t$. An infinitesimal symmetry characteristic for pKdV has to satisfy the following equation
\begin{equation}
\eta_t-3u_x\eta_x-\textstyle{\frac14}\eta_{xxx}=0.\label{pKdV1}
\end{equation}
The solution of (\ref{pKdV1}) for $\eta$ in the form (\ref{etapk}) contains the following characteristics
\begin{align}
&\eta_1=1,&\eta_7&=(\psi^{(2)})^2,\nonumber\\
&\eta_2=u_x,&\eta_8&=\psi^{(1)}\psi^{(2)},\nonumber\\
&\eta_3=u_t,&\eta_9&=\psi^{(1)}\psi^{(1)}_{\lambda},\label{etak}\\
&\eta_4=x+3tu_x,&\eta_{10}&=\psi^{(2)}\psi^{(2)}_{\lambda},\nonumber\\
&\eta_5=xu_x+3tu_t+u,&\eta_{11}&=\psi^{(1)}\psi^{(2)}_{\lambda}+\psi^{(2)}\psi^{(1)}_{\lambda},\nonumber\\
&\eta_6=(\psi^{(1)})^2,&\eta_{12}&=\psi^{(1)}\psi^{(2)}_{\lambda}-\psi^{(2)}\psi^{(1)}_{\lambda}-\textstyle{\frac{3}{2}}ct.\nonumber
\end{align}
\noindent{\bf Notes:} 
\begin{enumerate}
	\item The characteristics $\eta_{1-5}$ correspond to the standard point symmetries of pKdV. 
	\item $\eta_{6-12}$ are GS characteristics.
	\item  $\eta_{6-8}$ are square eigenfunctions. These symmetries were found in \cite{Dic,lou2012,OEVEL1998161} and described as GSs in \cite{RSG} in terms of the variable $v=\frac{\psi_x}{\psi}$. The characteristic $\eta_{12}$ was presented in \cite{RS7} in terms of the variable $z=\ln\left(\frac{\psi^{(1)}}{\psi^{(2)}}\right)$.
	\item The characteristics $\eta_{9-11}$ are the derivatives of $\eta_{6-8}$ with respect to $\lambda$, so they don't generate a new hierarchy of symmetries.
	\item  The GS characteristics for KdV can be obtained by differentiation of (\ref{etak}) with respect to $x$.
\end{enumerate}
\subsection{The Camassa-Holm equation}
The Camassa-Holm equation (CH) \cite{CH0} is
\begin{equation}
m_t+2u_xm+um_x=0,\qquad    m=u-u_{xx},\label{CM0}
\end{equation}
or equivalently 
\begin{equation}
u_t-u_{txx}+3uu_x-uu_{xxx}-2u_xu_{xx}=0.  \label{CM1} 
\end{equation}
Writing $u=v_x$ and integrating once, we obtain the potential Camassa-Holm equation (pCH) 
\begin{equation}
v_t-v_{txx}+\frac{3}{2}v_x^2-v_xv_{xxx}-\frac{1}{2}v_{xx}^2=0 \ . \label{CM} 
\end{equation}
The Lax pair for CH and pCH is
\begin{eqnarray}
\psi_{xx}&=&\left(\frac{1}{4}-\frac{1}{2\lambda}m\right)\psi,\label{LPCH1}\\
\psi_t&=&-(\lambda+v_x)\psi_x+\frac12 v_{xx}\psi,\label{LPCH2}
\end{eqnarray}
The coefficient of $\psi_x$ in (\ref{LPCH1}) is zero, therefore formula (\ref{plk}) is also true for pCH. From the consistency of (\ref{plk}) with (\ref{LPCH1},\ref{LPCH2}) we obtain that $c$ does not depend upon $t$.
 An infinitesimal symmetry characteristic for pCH has to satisfy the following equation
\begin{equation}
\eta_t-\eta_{txx}+3u_x\eta_x-v_{xxx}\eta_x-v_{x}\eta_{xxx}-v_{xx}\eta_{xx}=0.\label{pCH1}
\end{equation}
The solution of (\ref{pCH1}) for $\eta$ in the form (\ref{etapk}) contains the following linearly independent characteristics:
\begin{align}
&\eta_1=1,&\eta_7&=(\psi^{(2)})^2,\nonumber\\
&\eta_2=v_x,&\eta_8&=\psi^{(1)}\psi^{(1)}_{\lambda},\nonumber\\
&\eta_3=v_t,&\eta_9&=\psi^{(2)}\psi^{(2)}_{\lambda},\label{etaCH}\\
&\eta_4=tv_t+v,&\eta_{10}&=\psi^{(1)}\psi^{(2)}_{\lambda}+\psi^{(2)}\psi^{(1)}_{\lambda},\nonumber\\
&\eta_5=\psi^{(1)}\psi^{(2)},&\eta_{11}&=\lambda(\psi^{(1)}\psi^{(2)}_{\lambda}-\psi^{(2)}\psi^{(1)}_{\lambda})+ct(\lambda+v_x).\nonumber\\
&\eta_6=(\psi^{(1)})^2,&&\nonumber
\end{align}
\noindent{\bf Notes:} 
\begin{enumerate}
	\item The characteristics $\eta_{1-4}$ correspond to the standard point symmetries of pCH.
	\item $\eta_{5-11}$ are GS characteristics.
	\item $\eta_{5}$ was found in \cite{RS3} in terms of the variable $s=\frac{\psi_x}{\psi}$. $\eta_{5-7}$ are the square eigenfunctions. 
	\item The characteristics $\eta_{8-10}$ are derivatives of $\eta_{5-7}$ with respect to $\lambda$, so they don't generate a new hierarchy of symmetries. 
	\item  The generating symmetry characteristics for CH can be obtained by differentiation of (\ref{etaCH}) with respect to $x$.
\end{enumerate}
\subsection{The modified KdV equation}
The defocusing modified KdV equation (mKdV) is
\begin{equation}
  u_t - 6u^2u_x + u_{xxx} = 0 \ . \label{mKdV} 
\end{equation}  
The Lax pair for mKdV \cite{cn1, Sa0} is
\begin{eqnarray}
\psi_{xx}&=&\frac{u_x}{u}\psi_x+\left(u^2+\lambda \frac{u_x}{u}+\lambda^2\right)\psi,\label{LPmkdv1}\\
\psi_t&=&\left(\frac{2\lambda u_x-u_{xx}}{u}+2u^2-4\lambda^2\right)\psi_x-\frac{\lambda u_{xx}-2\lambda^2u_x}{u}\psi.\label{LPmkdv2}
\end{eqnarray}
The coefficient of $\psi_x$ is $\frac{u_x}{u}$, therefore from (\ref{pp1}) we get
\begin{equation}
\psi^{(2)}_x=\frac{\psi^{(1)}_x\psi^{(2)}+cu}{\psi^{(1)}}.\label{plmk}
\end{equation}
By checking the consistency with (\ref{LPmkdv1},\ref{LPmkdv2}) we obtain that $c$ does not depend upon $t$. 
An infinitesimal symmetry characteristic for mKdV has to satisfy the following equation
\begin{equation}
\eta_t-6u^2\eta_x-12uu_x\eta+\eta_{xxx}=0.\label{mKdV1}
\end{equation}
The solution of (\ref{mKdV1}) for $\eta$ in the form (\ref{etapk}) contains the following linearly independent characteristics (the result was simplified with the help of (\ref{plmk})):
\begin{align}
&\eta_1=u_x,&\eta_4&=\frac{(\lambda\psi^{(1)}+\psi^{(1)}_x)^2}{u^2}+(\psi^{(1)})^2,\nonumber\\
&\eta_2=u_t,&\eta_5&=\frac{(\lambda\psi^{(2)}+\psi^{(2)}_x)^2}{u^2}+(\psi^{(2)})^2,\label{etamk}\\
&\eta_3=xu_x+3tu_t+u,&\eta_{6}&=\frac{(\lambda\psi^{(1)}+\psi^{(1)}_x)(\lambda\psi^{(2)}+\psi^{(2)}_x)}{u^2}+\psi^{(1)}\psi^{(2)}.\nonumber
\end{align}
\noindent{\bf Notes:} 
\begin{enumerate}
	\item The characteristics $\eta_{1-3}$ correspond to the standard point symmetries of mKdV. 
	\item $\eta_{4-6}$ are generating symmetry characteristics. 
	\item In order to find additional GSs we need to include more variables in (\ref{etapk}). Computational difficulties don't allow us do this at this stage. However, with the help of intuition we managed to find a fourth GS
\[
\eta_{7}=\frac{(\lambda\psi_{\lambda}^{(1)}+\psi^{(1)}+\psi_{\lambda x}^{(1)})(\lambda\psi^{(2)}+\psi^{(2)}_x)}{u^2}+\psi^{(1)}_{\lambda}\psi^{(2)}+6ctu_x.
\]
\end{enumerate}
There is another Lax pair for mKdV \cite{Sch1,clarkson}:
\begin{eqnarray}
\phi_{xx}&=&\left(u^2+u_x+\lambda^2\right)\phi,\label{LPmkdv12}\\
\phi_t&=&3(u^2+u_x-\lambda^2)\phi_x-\phi_{xxx},\label{LPmkdv22}
\end{eqnarray}
This Lax pair is connected to (\ref{LPmkdv1},\ref{LPmkdv2}) via substitutions
\begin{equation}
\phi=\frac{\psi_x+(u+\lambda)\psi}{u},~~~\psi=\frac{1}{2\lambda}((u+\lambda)\phi-\phi_x).\label{psiphi}
\end{equation}
These substitutions can be derived from a zero curvature representation of mKdV \cite{Sch1}. With the help of (\ref{psiphi}) all GSs characteristics $\eta_{4-7}$ can be expressed via the variable $\phi$.
\subsection{The sine-Gordon equation}
The sine-Gordon equation (SG) is
\begin{equation}
u_{tx}=\sin(u).
\end{equation}
The Lax pair for SG \cite{AKNS,Kaup} is
\begin{eqnarray}
\psi_{xx}&=&\frac{u_{xx}}{u_x}\psi_x+\left(\lambda^2-\lambda \frac{u_{xx}}{u_x}-\frac{u_x^2}{4}\right)\psi,\label{LPSG1}\\
\psi_t&=&-\frac{\sin u}{2\lambda u_x}\psi_x+\left(\frac{\cos u}{4\lambda}+\frac{\sin u}{2u_x}\right)\psi.\label{LPSG2}
\end{eqnarray}
The coefficient of $\psi_x$ in (\ref{LPSG1}) is $\frac{u_{xx}}{u_x}$, therefore from (\ref{pp1}) we get
\begin{equation}
\psi^{(2)}_x=\frac{\psi^{(1)}_x\psi^{(2)}+cu_x}{\psi^{(1)}}.\label{plSG}
\end{equation}
By checking the consistency with (\ref{LPSG1},\ref{LPSG2}) we obtain that $c$ does not depend upon $t$. An infinitesimal symmetry characteristic for SG has to satisfy the following equation
\begin{equation}
\eta_{tx}-\cos(u)\eta=0.\label{SG1}
\end{equation}
The solution of (\ref{SG1}) for $\eta$ in the form (\ref{etapk}) contains the following linearly independent characteristics (the result was simplified with the help of (\ref{plSG})):
\begin{align}
&\eta_1=u_x,&\eta_4&=\frac{4(\lambda\psi^{(1)}-\psi^{(1)}_x)^2}{u_x^2}+(\psi^{(1)})^2,\nonumber\\
&\eta_2=u_t,&\eta_5&=\frac{4(\lambda\psi^{(2)}-\psi^{(2)}_x)^2}{u_x^2}+(\psi^{(2)})^2,\label{etasg}\\
&\eta_3=xu_x-tu_t,&\eta_{6}&=\frac{4(\lambda\psi^{(1)}-\psi^{(1)}_x)(\lambda\psi^{(2)}-\psi^{(2)}_x)}{u_x^2}+\psi^{(1)}\psi^{(2)}.\nonumber
\end{align}
\noindent{\bf Notes:} 
\begin{enumerate}
	\item The characteristics $\eta_{1-3}$ correspond to the standard point symmetries of SG. 
	\item $\eta_{4-6}$ are GS characteristics. 
	\item In order to find additional GSs we need to include more variables in (\ref{etapk}). Computational difficulties don't allow us do this at this stage. However, with the help of intuition we managed to find a fourth GS
\[
\eta_{7}=\frac{4(\lambda\psi_{\lambda}^{(1)}+\psi^{(1)}-\psi_{\lambda x}^{(1)})(\lambda\psi^{(2)}-\psi^{(2)}_x)}{u^2}+\psi^{(1)}_{\lambda}\psi^{(2)}+\frac{c}{\lambda}tu_t.
\]
\item Remarkably, all GSs for SG are very similar to GSs for mKdV.
\end{enumerate}
There is another Lax pair for SG \cite{Nucci}:
\begin{eqnarray}
\phi_{xx}&=&-i u\phi_x+\lambda^2\phi,\label{LPSG12}\\
\phi_t&=&\frac{e^{i u}}{4\lambda^2}\phi_x-\frac{i}{2}u_t\phi,\label{LPSG22}
\end{eqnarray}
This Lax pair is connected to (\ref{LPSG1},\ref{LPSG2}) via substitutions
\begin{equation}
\phi=\frac{(2\psi_x+(i u_x-2\lambda)\psi)}{2u_xe^{\frac{i u}{2}}},~~~\psi=-\frac{i}{\lambda}(\phi_x+\lambda\phi)e^{\frac{i u}{2}}.\label{psiphi1}
\end{equation}
With the help of (\ref{psiphi1}) all GSs characteristics $\eta_{4-7}$ can be expressed via variable $\phi$.

The SG equation can be brought to rational form by the transformation $u=2i\ln(z)$: 
\begin{equation}
zz_{tx}-z_xz_t=\textstyle{\frac{1}{4}}(z^4-1).\label{sggq}
\end{equation}
The Lax pair (\ref{LPSG12},\ref{LPSG22}) transforms to 
\begin{eqnarray}
\phi_{xx}&=&2\frac{z_x}{z}\phi_x+\lambda^2\phi,\label{LPSG13}\\
\phi_t&=&\frac{\phi_x}{4\lambda^2z^2}+\frac{z_t}{z}\phi.\label{LPSG23}
\end{eqnarray}
The coefficient of $\phi_x$ in (\ref{LPSG13}) is $\frac{2z_{x}}{z}$, therefore from (\ref{pp1}) we get
\begin{equation}
\phi^{(2)}_x=\frac{\phi^{(1)}_x\phi^{(2)}+cz^2}{\phi^{(1)}}.\label{plSG33}
\end{equation}
By checking the consistency with (\ref{LPSG13},\ref{LPSG23}) we obtain that $c$ does not depend upon $t$. 
An infinitesimal symmetry characteristic for (\ref{sggq}) has to satisfy the following equation
\begin{equation}
z\eta_{tx}+\eta z_{tx}-\eta_x z_t-z_x\eta_t=z^3\eta.
\end{equation}
The equation (\ref{sggq}) admits the following particularly simple GS characteristics:
\begin{align*}
&\eta_1=z_x,&\eta_5&=\frac{\phi^{(2)}\phi^{(2)}_x}{z},\\
&\eta_2=z_t,&\eta_6&=\frac{\phi^{(1)}\phi^{(2)}_x+\phi^{(2)}\phi^{(1)}_x}{z},\\
&\eta_3=xz_x-tz_t,&\eta_7&=\frac{\phi^{(1)}_{\lambda}\phi^{(2)}_x+\phi^{(1)}_{\lambda x}\phi^{(2)}}{z}+\frac{c}{\lambda}\left(\frac{z}{2}+tz_t\right).\\
&\eta_4=\frac{\phi^{(1)}\phi^{(1)}_x}{z},& &
\end{align*}
\noindent{\bf Notes:} 
\begin{enumerate}
	\item The characteristics $\eta_{1-3}$ correspond to the standard point symmetries. 
	\item $\eta_{4-7}$ are GS characteristics. 
	\item These symmetries can also be obtained from (\ref{etasg}) via (\ref{psiphi1}) and $u=2i\ln(z)$.
	\item $\eta_{5}$ was found in \cite{RSG} in terms of the variables $z^{(1)}=\frac{\phi^{(1)}_x}{\phi^{(1)}}$ and $z^{(2)}=\frac{\phi^{(2)}_x}{\phi^{(2)}}$.
\end{enumerate}
\section{Generating symmetries for equations with a third order Lax pair}
The general form of (\ref{LP1}) in the third order Lax pair is 
\begin{equation}
\psi_{xxx}=F_1\psi_{xx}+F_2\psi_x+F_3\psi.\label{LPt3}
\end{equation}
Here $F_1, F_2, F_3$ are functions, which depend upon the parameter $\lambda$, variables $t, x, u$ and derivatives of $u$.

\noindent{\bf Notes:} 
\begin{enumerate}
	\item  The third order Lax pair has three linearly independent solutions: $\psi^{(1)}, \psi^{(2)}$ and $\psi^{(3)}$.
	\item  The three solutions $\psi^{(1)}, \psi^{(2)}, \psi^{(3)}$ satisfy
\begin{multline}
W(\psi^{(1)},\psi^{(2)},\psi^{(3)})=(\psi^{(1)}\psi^{(2)}_x-\psi^{(1)}_x\psi^{(2)})\psi^{(3)}_{xx}-\\(\psi^{(1)}\psi^{(3)}_x-\psi^{(1)}_x\psi^{(3)})\psi^{(2)}_{xx}+(\psi^{(2)}\psi^{(3)}_x-\psi^{(2)}_x\psi^{(3)})\psi^{(1)}_{xx}=ce^{\int F_1 dx},\label{ppp1}
\end{multline}
where $W$ is the Wronskian and $c$ is a constant of integration independent of $x$. The dependence of $c$ upon $t$ can be determined from the consistency of (\ref{ppp1}) and the Lax pair. We ssume that $\int F_1 dx$ is local.
\end{enumerate}

\subsection{The Boussinesq equation}
The Boussinesq equation is
\begin{eqnarray*}
  u_t &=&  \left(-2v-u_x \right)_x \ ,  \\
  v_t &=&  \left(v_x + {\textstyle{\frac23}} u_{xx} -u^2 \right)_x  \ . 
\end{eqnarray*}
This is the two component form. By eliminating $v$ we obtain the scalar form of the Boussinesq equation: 
\[
u_{tt} = -{\textstyle{\frac13}} u_{xxxx} + (2u^2)_{xx}  \ .
\]
The Boussinesq equation is connected to the potential Boussinesq equation (pBE) via $u=f_x$, $v=w_x$
\begin{eqnarray}
  f_t &=&  \left(-2 w - f_x \right)_x \ , \label{ueq}\\
  w_t &=&  w_{xx} + {\textstyle{\frac23}} f_{xxx} - f_x^2  \ .\label{veq}
\end{eqnarray}
By eliminating $w$ from (\ref{ueq},\ref{veq}) we obtain the scalar form of pBE: 
\begin{equation}
f_{tt} = -{\textstyle{\frac13}} f_{xxxx} + 4f_xf_{xx}  \ .
\label{usceq}\end{equation}
The Lax pair for Boussinesq and potential Boussinesq equations \cite{abl1} is
\begin{eqnarray}
\psi_{xxx}&=&3f_x\psi_x+(\lambda-3w_x)\psi,\label{LPB1}\\
\psi_t&=&\psi_{xx}-2f_x\psi.\label{LPB2}
\end{eqnarray}
The coefficient of $\psi_{xx}$ is zero, therefore from (\ref{ppp1}) we get
\begin{equation}
\psi^{(3)}_{xx}=\frac{(\psi^{(1)}\psi^{(3)}_x-\psi^{(1)}_x\psi^{(3)})\psi^{(2)}_{xx}-(\psi^{(2)}\psi^{(3)}_x-\psi^{(2)}_x\psi^{(3)})\psi^{(1)}_{xx}+c}{\psi^{(1)}\psi^{(2)}_x-\psi^{(1)}_x\psi^{(2)}}.\label{plb}
\end{equation}
By checking the consistency with (\ref{LPB1},\ref{LPB2}) we obtain, that $c$ does not depend upon $t$. An infinitesimal symmetry characteristic for pBE has to satisfy the following equation
\begin{equation}
\eta_{tt}-4f_x\eta_{xx}-4f_{xx}\eta_{x}+\textstyle{\frac13}\eta_{xxxx}=0.\label{bus1}
\end{equation}
In order to find generating symmetries for pBE we take
\begin{equation}
\eta=\eta(x,t,f,f_x,f_t,\psi^{(1)}, \psi^{(2)}, \psi^{(3)},\psi^{(1)}_x,\psi^{(2)}_x,\psi^{(3)}_x, \psi^{(1)}_{\lambda}, \psi^{(2)}_{\lambda},\psi^{(3)}_{\lambda}).\label{etapb}
\end{equation}
The solution of (\ref{bus1}) for $\eta$ in the form (\ref{etapb}) gives the following linearly independent characteristics:
\begin{align}
&\eta_1=1,&\eta_9&=\psi^{(2)}(\psi^{(2)}\psi^{(3)}_x-\psi^{(2)}_x\psi^{(3)}),\nonumber\\
&\eta_2=t,&\eta_{10}&=\psi^{(3)}(\psi^{(1)}\psi^{(3)}_x-\psi^{(1)}_x\psi^{(3)}),\nonumber\\
&\eta_3=f_x,&\eta_{11}&=\psi^{(3)}(\psi^{(2)}\psi^{(3)}_x-\psi^{(2)}_x\psi^{(3)}),\nonumber\\
&\eta_4=f_t,&\eta_{12}&=\psi^{(1)}(\psi^{(2)}\psi^{(1)}_x-\psi^{(2)}_x\psi^{(1)}),\label{pbeeta}\\
&\eta_5=xf_x+2tf_t+f,&\eta_{13}&=\psi^{(2)}(\psi^{(1)}\psi^{(2)}_x-\psi^{(1)}_x\psi^{(2)}),\nonumber\\
&\eta_6=\psi^{(2)}(\psi^{(1)}\psi^{(3)}_x-\psi^{(1)}_x\psi^{(3)}),&\eta_{14}&=\psi^{(3)}_{\lambda}(\psi^{(1)}\psi^{(2)}_x-\psi^{(1)}_x\psi^{(2)})\nonumber\\
&\eta_7=\psi^{(1)}(\psi^{(2)}\psi^{(3)}_x-\psi^{(2)}_x\psi^{(3)}),& &+\psi^{(2)}_{\lambda}(\psi^{(3)}\psi^{(1)}_x-\psi^{(3)}_x\psi^{(1)})\nonumber\\
&\eta_8=\psi^{(1)}(\psi^{(1)}\psi^{(3)}_x-\psi^{(1)}_x\psi^{(3)}),& &+\psi^{(1)}_{\lambda}(\psi^{(2)}\psi^{(3)}_x-\psi^{(2)}_x\psi^{(3)}).\nonumber
\end{align}
\noindent{\bf Notes:}
\begin{enumerate}
	\item The characteristics $\eta_{1-5}$ correspond to the standard point symmetries of pBE. 
	\item $\eta_{6-14}$ are GS characteristics.
	\item $\eta_{6-7}$ were found in \cite{RS4} in terms of the variable $s=\frac{\psi_x}{\psi}$.
	\item The characteristics $\eta_{8-13}$ can be obtained from $\eta_{6-7}$ by taking $\psi^{(i)}=\psi^{(j)}$ for some $i\neq j$. This fact was also noted in \cite{RS4}.
	\item It would be good to add $\psi^{(1)}_{xx}, \psi^{(2)}_{xx}$ and $\psi^{(3)}_{xx}$ in $\eta$ in order to find more GSs. Computational difficulties don't allow us do this at this stage.
	\item  The GS characteristics for the Boussinesq equation can be obtained by differentiation of (\ref{pbeeta}) with respect to $x$.
\end{enumerate}

\subsection{The associated Degasperis-Procesi equation}
The associated Degasperis-Procesi equation (aDP) is  
\begin{equation}   
  f_{xxt} - \textstyle{\frac34} \frac{f_{xt}^2}{f_t} +  3( 1 - f_x f_t )  = 0  \ .    \label{myDP}
\end{equation}
The integrability properties of this equation were presented in \cite{RS5}. The Lax pair was presented there, as well:
\begin{eqnarray}
  \psi_{xxx} &=&  3 f_x \psi_x + \left(  \frac32 f_{xx} + \theta \right) \psi \ ,  \label{psixxxD} \\
  \psi_t    &=& \frac1{\theta}\left(  f_t \psi_{xx} - \frac12 f_{tx} \psi_{x}  +
   \left( \frac18 \frac{f_{xt}^2}{f_t} - \frac32 f_x f_t \right) \psi 
       \right)
     \ .  \label{psitD}
\end{eqnarray}
The coefficient of $\psi_{xx}$ is zero, therefore (\ref{plb}) is also true. From the consistency of (\ref{plb}) with (\ref{psixxxD},\ref{psitD}) we obtain that $c=c_1e^{\frac{3t}{2\lambda}}$ where $c_1$ does not depend on $t$.
An infinitesimal symmetry characteristic for aDP has to satisfy the following equation
\begin{equation}
\eta_{xxt}-\frac32\eta_{xt}\frac{f_{xt}}{f_t}+\frac{3}{4}\eta_t \frac{f_{xt}^2}{f_t^2}-3f_{t}\eta_{x}-3\eta_{t}f_{x}=0.\label{aDPl1}
\end{equation} 
In order to find generating symmetries for aDP we take
\begin{equation}
\eta=\eta(x,t,f,f_x,f_t,\psi^{(1)}, \psi^{(2)},\psi^{(3)},\psi^{(1)}_x,\psi^{(2)}_x,\psi^{(3)}_x).\label{etaDP}
\end{equation}
We did not take $\eta$ in the form (\ref{etapb}), because of the computational difficulties. After removing variables $\psi^{(1)}_{\lambda}, \psi^{(2)}_{\lambda},\psi^{(3)}_{\lambda}$ calculations have been completed. 
The solution of (\ref{aDPl1}) for $\eta$ in the form (\ref{etaDP}) gives the following linearly independent characteristics:
\begin{align}
&\eta_1=1,&\eta_7&=\psi^{(2)}(\psi^{(2)}\psi^{(3)}_x-\psi^{(2)}_x\psi^{(3)}),\nonumber\\
&\eta_2=f_x,&\eta_{8}&=\psi^{(3)}(\psi^{(2)}\psi^{(3)}_x-\psi^{(2)}_x\psi^{(3)}),\nonumber\\
&\eta_3=f_t,&\eta_{9}&=\psi^{(1)}(\psi^{(2)}\psi^{(1)}_x-\psi^{(2)}_x\psi^{(1)}),\label{aDPeta}\\
&\eta_4=xf_x-3tf_t+f,&\eta_{10}&=\psi^{(2)}(\psi^{(1)}\psi^{(2)}_x-\psi^{(1)}_x\psi^{(2)}),\nonumber\\
&\eta_5=\psi^{(2)}(\psi^{(1)}\psi^{(3)}_x-\psi^{(1)}_x\psi^{(3)}),&\eta_{11}&=\psi^{(1)}(\psi^{(1)}\psi^{(3)}_x-\psi^{(1)}_x\psi^{(3)}),\nonumber\\
&\eta_6=\psi^{(1)}(\psi^{(2)}\psi^{(3)}_x-\psi^{(2)}_x\psi^{(3)}),& \eta_{12}&=\psi^{(3)}(\psi^{(1)}\psi^{(3)}_x-\psi^{(1)}_x\psi^{(3)}).\nonumber
\end{align}
\noindent{\bf Notes:}
\begin{enumerate}
	\item $\eta_{5-12}$ are GS characteristics.
	\item $\eta_{5-6}$ were found in \cite{RS5} in terms of the variable $s=\frac{\psi_x}{\psi}$.
	\item The characteristics $\eta_{7-12}$ can be obtained from $\eta_{5-6}$ by taking $\psi^{(i)}=\psi^{(j)}$ for some $i\neq j$. 
	\item In order to find additional GSs we need to include more variables in (\ref{etaDP}). Computational difficulties don't allow us to do it at this stage. However, with the help of intuition we manage to find one more GS
\begin{multline}
\eta_{13}=\left(\psi^{(3)}_{\lambda}(\psi^{(1)}\psi^{(2)}_x-\psi^{(1)}_x\psi^{(2)})+\psi^{(2)}_{\lambda}(\psi^{(3)}\psi^{(1)}_x-\psi^{(3)}_x\psi^{(1)})\right.\\\left.+\psi^{(1)}_{\lambda}(\psi^{(2)}\psi^{(3)}_x-\psi^{(2)}_x\psi^{(3)})\right)e^{-\frac{3t}{2\lambda}}+\frac{c_1t}{\lambda^2}f_t.
\end{multline}
\end{enumerate}
\subsection{The associated Novikov equation}
The associated Novikov equation (aN) is
\begin{equation}
f_{xxt}-3(f_xf_t-1)=0\ .   \label{aN} 
\end{equation}
The integrability properties of this equation were presented in \cite{RS6}. The point symmetries were found there, as well:
\begin{eqnarray}
  \eta_1 &=&1 \ , \nonumber\\
  \eta_2 &=&f_x \ , \label{syman}\\
  \eta_3 &=&f_t \ , \nonumber\\
  \eta_4 &=&xf_x-3tf_t+f\ .\nonumber
\end{eqnarray}
The Lax pair for aN is 
\begin{align}
\psi_{xxx}&=3f_x\psi_x+\lambda\psi,&\label{psixxx}\\
\psi_t&=\frac{1}{\lambda}(f_t\psi_{xx}-f_{xt}\psi_x).\label{psit}&
\end{align}
The coefficient of $\psi_{xx}$ is zero, therefore (\ref{plb}) is also true. From the consistency of (\ref{plb}) with (\ref{psixxx},\ref{psit}) we obtain that $c=c_1e^{\frac{6t}{\lambda}}$  where $c_1$ does not depend on $t$.
An infinitesimal symmetry characteristic for aN has to satisfy the following equation
\begin{equation}
\eta_{xxt}-3f_x\eta_{t}-3f_{t}\eta_{x}=0.\label{aNl1}
\end{equation}
In order to find GSs for aN we take
\begin{equation}
\eta=\eta(t,\psi^{(1)}, \psi^{(2)},\psi^{(3)},\psi^{(1)}_x,\psi^{(2)}_x,\psi^{(3)}_x, \psi^{(1)}_{xx}, \psi^{(2)}_{xx}).\label{etan}
\end{equation}
The $\eta$ doesn't depend upon $\psi^{(3)}_{xx}$, since it can be eliminated with the help of (\ref{plb}). This is the only form in which we could find GSs. Any additional variable causes incompletable calculations.
The solution of (\ref{aNl1}) for $\eta$ in the form (\ref{etan}) gives the following linearly independent characteristics (the result was simplified with the help of (\ref{plb})):
\begin{eqnarray*}
  \eta_1 &=&1 \ , \\
\eta_5&=&\left(\lambda\psi^{(1)}(\psi^{(2)}\psi^{(3)}_x-\psi^{(2)}_x\psi^{(3)})+\psi^{(1)}_x(\psi^{(2)}_x\psi^{(3)}_{xx}-\psi^{(2)}_{xx}\psi^{(3)}_x)\right)e^{-\frac{6t}{\lambda}},\\
\eta_6&=&\left(\lambda\psi^{(3)}(\psi^{(1)}\psi^{(2)}_x-\psi^{(1)}_x\psi^{(2)})+\psi^{(3)}_x(\psi^{(1)}_x\psi^{(2)}_{xx}-\psi^{(1)}_{xx}\psi^{(2)}_x)\right)e^{-\frac{6t}{\lambda}},\\
\eta_7&=&\left(\lambda\psi^{(3)}(\psi^{(2)}\psi^{(3)}_x-\psi^{(2)}_x\psi^{(3)})+\psi^{(3)}_x(\psi^{(2)}_x\psi^{(3)}_{xx}-\psi^{(2)}_{xx}\psi^{(3)}_x)\right)e^{-\frac{6t}{\lambda}},\\
\eta_8&=&\left(\lambda\psi^{(3)}(\psi^{(1)}\psi^{(3)}_x-\psi^{(1)}_x\psi^{(3)})+\psi^{(3)}_x(\psi^{(1)}_x\psi^{(3)}_{xx}-\psi^{(1)}_{xx}\psi^{(3)}_x)\right)e^{-\frac{6t}{\lambda}},\\
\eta_9&=&\left(\lambda\psi^{(2)}(\psi^{(2)}\psi^{(3)}_x-\psi^{(2)}_x\psi^{(3)})+\psi^{(2)}_x(\psi^{(2)}_x\psi^{(3)}_{xx}-\psi^{(2)}_{xx}\psi^{(3)}_x)\right)e^{-\frac{6t}{\lambda}},\\
\eta_{10}&=&\left(\lambda\psi^{(1)}(\psi^{(1)}\psi^{(3)}_x-\psi^{(1)}_x\psi^{(3)})+\psi^{(1)}_x(\psi^{(1)}_x\psi^{(3)}_{xx}-\psi^{(1)}_{xx}\psi^{(3)}_x)\right)e^{-\frac{6t}{\lambda}},\\
\eta_{11}&=&\left(\lambda\psi^{(2)}(\psi^{(1)}\psi^{(2)}_x-\psi^{(1)}_x\psi^{(2)})+\psi^{(2)}_x(\psi^{(1)}_x\psi^{(2)}_{xx}-\psi^{(1)}_{xx}\psi^{(2)}_x)\right)e^{-\frac{6t}{\lambda}},\\
\eta_{12}&=&\left(\lambda\psi^{(1)}(\psi^{(1)}\psi^{(2)}_x-\psi^{(1)}_x\psi^{(2)})+\psi^{(1)}_x(\psi^{(1)}_x\psi^{(2)}_{xx}-\psi^{(1)}_{xx}\psi^{(2)}_x)\right)e^{-\frac{6t}{\lambda}}.\\
\end{eqnarray*}
\noindent{\bf Notes:}
\begin{enumerate}
  \item $\eta_1$ is the point symmetry shown in (\ref{syman}). Other point symmetries can not be obtained from $\eta$ in the form (\ref{etan}), that is why they are not listed here. 
	\item $\eta_{5-12}$ are GS characteristics.
	\item $\eta_{5-6}$ were found in \cite{RS6} in terms of the variable $v=\frac{\psi_x}{\psi}$.
	\item The characteristics $\eta_{7-12}$ can be obtained from $\eta_{5-6}$ by taking $\psi^{(i)}=\psi^{(j)}$ for some $i\neq j$. 
	\item In order to find additional GSs we need to include more variables in (\ref{etan}). Computational difficulties don't allow us to do it at this stage. However, with the help of intuition we manage to find one more GS
\begin{eqnarray*}
\eta_{13}&=&\left(\lambda\psi^{(1)}_{\lambda}(\psi^{(2)}\psi^{(3)}_x-\psi^{(2)}_x\psi^{(3)})+\psi^{(1)}_{\lambda x}(\psi^{(2)}_x\psi^{(3)}_{xx}-\psi^{(2)}_{xx}\psi^{(3)}_x)\right)e^{-\frac{6t}{\lambda}}\\
&+&\left(\lambda\psi^{(3)}_{\lambda}(\psi^{(1)}\psi^{(2)}_x-\psi^{(1)}_x\psi^{(2)})+\psi^{(3)}_{\lambda x}(\psi^{(1)}_x\psi^{(2)}_{xx}-\psi^{(1)}_{xx}\psi^{(2)}_x)\right)e^{-\frac{6t}{\lambda}}\\
&+&\left(\lambda\psi^{(2)}_{\lambda}(\psi^{(3)}\psi^{(1)}_x-\psi^{(3)}_x\psi^{(1)})+\psi^{(2)}_{\lambda x}(\psi^{(3)}_x\psi^{(1)}_{xx}-\psi^{(3)}_{xx}\psi^{(1)}_x)\right)e^{-\frac{6t}{\lambda}}+\frac{2c_1}{\lambda}tf_t.
\end{eqnarray*}
\end{enumerate}
\section{Conclusion}
In this paper we present the method for computation of GSs. On the one hand, it is a simple upgrade of the standard symmetry method. On the other hand, inclusion of $\psi$ and $\lambda$ seems to refine symmetry analysis to a finite problem. The detailed implementation of GSM is described in the Maple program. As one can see, the idea of the method is very simple, but effective. The main obstacle of GSM is  computational difficulties, which involve the solution of large systems of PDEs.

GSM was applied to seven integrable equations: KdV, CH, mKdV, sine-Gordon, Boussinesq, associated Degasperis-Procesi and associated Novikov. All of them are integrable and admit a Lax pair. It is remarkable, that for all the equations GSs are found. In the case of equations with the second order Lax pair four essential GSs were found. From the results of \cite{RS7} it follows that this is, probably, a complete list of GSs. In the case of equations with the third order Lax pair nine essential GSs were found.
 
Directions for further research:
\begin{itemize}
	\item Discovering GSs for more equations, including higher dimensional equations, for example KP and the Calogero--Bogoyavlensky--Schiff equation \cite{S2}.
	\item Expansion of obtained GSs into hierarchies of symmetries. Especially to find expansions of GSs which give scaling symmetries for Camassa-Holm and associated Degasperis-Procesi equations.
\item Using GSs to find new solutions of the equations considered in this paper.
	\item Determining the commutator relations of found GSs. For this we need to compute the action of GSs on $\psi$. At this moment it is not clear how to do it systematically.
	\item Understanding the connection between the number of GSs and the order of a Lax pair.
	\item Classification for different types of GSs.
	\item Classification of equations with respect to GSs.
\end{itemize}
\noindent

{\bf \large Acknowledgments}

I would like to thank Prof. Jeremy Schiff for encouraging this line of research and for significant comments.
\section{Appendix}
Here we present the Maple program, which computes the generating symmetries for pKdV. We use the following notations:
\begin{itemize}
	\item $Q$ denotes a symmetry characteristic.
	\item pKdV has a second order Lax pair, which admits two linearly independent solutions. We denote them $w(t,x,\lambda)$ and $v(t,x,\lambda)$.
\end{itemize}
{\setstretch{1.5}

\begin{maplelatex}
\mapleinline{inert}{1d}{
with(DEtools):
with(StringTools):
}{}
\end{maplelatex}
First, we introduce the following notation:

\begin{maplelatex}
\mapleinline{inert}{1d}{
tru:=seq(seq(ifelse(i+j=0,u(t, x),diff(u(t,x),x\$i,t\$j))=convert(Join (["u",Repeat("t",j),Repeat("x",i)],""),name),i=0..4),j=0..2);
trv:=seq(seq(seq(ifelse(i+j+k=0,v(t,x,lambda),diff(v(t,x,lambda),x\$i ,t\$j,lambda\$k))=convert(Join(["v",Repeat("l",k),Repeat("t",j),Repeat("x",i )],""),name),i=0..4),j=0..1),k=0..1):
trw:=seq(seq(seq(ifelse(i+j+k=0,w(t,x,lambda),diff(w(t,x,lambda),x\$i ,t\$j,lambda\$k))=convert(Join(["w",Repeat("l",k),Repeat("t",j),Repeat("x",i )],""),name),i=0..4),j=0..1),k=0..1):
tr:=\{tru,trv,trw,C(lambda)=c,diff(C(lambda),lambda)=cl\}:
}{}
\end{maplelatex}
Here $vx, vt, vl$ denote the derivatives of $v$ with respect to $x, t, \lambda$. We use the same notation for variables $u,w$.
Let us define the infinitesimal generator with prolongations:
 
\begin{maplelatex}
\mapleinline{inert}{1d}{
eta:=Q(t,x,u(t,x),diff(u(t,x),x),diff(u(t,x),t),v(t,x,lambda),diff(v (t,x,lambda),x),w(t,x,lambda),diff(v(t,x,lambda),lambda),diff(w(t,x,lambda ),lambda)):
X:=(A)->eta*diff(A,u)+diff(eta,x)*diff(A,ux)+diff(eta,t)*diff(A,ut)+ diff(eta,x,x)*diff(A,uxx)+diff(eta,x,x,x)*diff(A,uxxx):
}{}
\end{maplelatex}
The determining equation obtained by the action of the infinitesimal generator on pKdV is:

\begin{maplelatex}
\mapleinline{inert}{1d}{
R:=X(ut-1/4*uxxx-3/2*ux^2):
R1:=(subs(tr,R)):
}{}
\end{maplelatex}
We express the higher derivatives of the Lax pair, KdV and condition (\ref{plk}) in order to simplify the determining equation:

\begin{maplelatex}
\mapleinline{inert}{1d}{
LPV1:=diff(v(t,x,lambda),x,x)=(lambda-2*diff(u(t,x),x))*v(t,x,lambda):
LPV2:=diff(v(t,x,lambda),t)=(lambda+diff(u(t,x),x))*diff(v(t,x,lambda ),x)-1/2*diff(u(t,x),x,x)*v(t,x,lambda):
LPW1:=subs(v=w,LPV1):
LPW2:=subs(v=w,LPV2):
eq:=diff(u(t,x),x,x,x)=-6*diff(u(t,x),x)^2+4*diff(u(t,x),t):
W:=diff(w(t,x,lambda),x)=(diff(v(t,x,lambda),x)*w(t,x,lambda)+C(lambd a))/v(t,x,lambda):
}{}
\end{maplelatex}
\begin{maplelatex}
\mapleinline{inert}{1d}{
tr1:=subs(tr,{eq,diff(eq,x),diff(eq,t),LPV1,diff(LPV1,x),diff(LPV1,x, x),diff(LPV1,lambda),diff(LPV1,lambda,x),LPV2,diff(LPV2,x),diff(LPV2,lambd a),LPW1,diff(LPW1,x),diff(LPW1,x,x),diff(LPW1,lambda),diff(LPW1,lambda,x), LPW2,diff(LPW2,x),diff(LPW2,lambda),W,diff(W,lambda)}):
R2:=numer(factor(subs(tr1,subs(tr1,subs(tr1,R1))))):
}{}
\end{maplelatex}
The splitting of the determining equation into system is:

\begin{maplelatex}
\mapleinline{inert}{1d}{
sys:={coeffs(R2,{uxx,utx,utxx,utt,vlx})}:
}{}
\end{maplelatex}
The solution of the obtained system gives symmetry characteristics:

\begin{maplelatex}
\mapleinline{inert}{1d}{
sol:=rifsimp(convert(sys, diff));
pdsolve(sol['Solved']);
}{}
\end{maplelatex}
}
\bibliographystyle{acm}
    \bibliography{my1}
\end{document}